\documentclass[twocolumn,showpacs,preprintnumbers,amsmath,amssymb]{revtex4-1}
\usepackage{graphicx}
\usepackage{dcolumn}
\usepackage{bm}

\newcommand{\bef}{\begin{figure}}
\newcommand{\eef}{\end{figure}}

\newcommand{\be}{\begin{equation}}
\newcommand{\ee}{\end{equation}}
\newcommand{\bea}{\begin{eqnarray}}
\newcommand{\eea}{\end{eqnarray}}
\begin{document}

\title{Comparison of results from a 2+1D relativistic
  viscous hydrodynamic model to elliptic and hexadecapole flow of charged hadrons measured 
  in Au-Au collisions at $\sqrt{s_{\rm {NN}}}$ = 200 GeV}

\author{Victor Roy$^{1}$, A.K. Chaudhuri$^{1}$, and Bedangadas Mohanty$^{1,2}$}
\affiliation{$^1$Variable Energy Cyclotron Centre, Kolkata 700064, India. \\ and \\$^2$ School of Physical Sciences, National Institute of Science Education and Research, Bhubaneswar 751005, India.}


\date{\today}
\begin{abstract}

Simulated results from a 2+1D relativistic viscous hydrodynamic model
have been compared to the experimental data on the centrality dependence  
of invariant yield, elliptic flow ($v_{2}$), and hexadecapole flow ($v_{4}$)  as 
a function of transverse momentum ($p_{T}$) of charged hadrons in Au-Au  
collisions at $\sqrt{s_{\rm {NN}}}$ = 200 GeV. Results from two types of initial 
transverse energy density profile, one based on the Glauber model and
other based on Color-Glass-Condensate~(CGC) are presented.  We observe no
difference in the simulated results on the invariant yield of charged hadrons
for the calculations with different initial conditions. 
The comparison
to the experimental data on invariant yield of charged hadrons supports a shear viscosity to entropy density
ratio ($\eta/s$) between 0 to 0.12 for the 0-10\% to 40-50\%
collision centralities.  The simulated $v_{2}(p_{T})$ is found to be
higher for a fluid with CGC based initial condition compared to
Glauber based initial condition for a given collision centrality. 
Consequently the Glauber
based calculations when compared to the experimental data requires a
lower value of $\eta/s$ relative to CGC based calculations. In
addition, a centrality dependence of the estimated $\eta/s$ is observed
from the $v_{2}(p_{T})$ study.
The $v_{4}(p_{T})$ for the collision centralities
0-10\% to 40-50\% supports a $\eta/s$ value between 0 - 0.08 for a CGC
based initial condition. While simulated results using the Glauber
based initial condition for the ideal fluid evolution under estimates the $v_{4}(p_{T})$ for
collision centralities 0-10\% to 30-40\%.  

\end{abstract}
\pacs{25.75.Ld}
\maketitle

\section{INTRODUCTION}

Heavy-ion collisions at the Relativistic Heavy Ion Collider (RHIC) have provided evidence
for the formation of a hot and dense QCD matter~\cite{Arsene:2004fa,Back:2004je,Adams:2005dq,Adcox:2004mh,Gyulassy:2004zy}. 
This presents an unique opportunity to study
the transport properties, like shear viscosity to entropy density ratio ($\eta/s$), of
the QCD matter. There are two main theoretical approaches to estimate the value of $\eta/s$
from the experimental data. One based on a microscopic approach as in
transport theory ~\cite{Xu:2011fe,Greco:2011zz,Alver:2010dn,Demir:2008zz,Bhalerao:2005mm} and 
other related to a macroscopic approach through relativistic viscous
hydrodynamic calculations~\cite{Niemi:2011ix,Shen:2010uy,Heinz:2009cv,Chaudhuri:2009uk,Bozek:2009dw,Schenke:2010rr,Luzum:2008cw,Luzum:2009sb}.
In this work, we will compare our results from a relativistic 2+1 dimension viscous hydrodynamics
to recent high statistics experimental data on elliptic ($v_{2}$) and hexadecapole ($v_{4}$) flow 
of charged hadrons measured by the PHENIX Collaboration~\cite{Adare:2010ux,Adler:2003au}. The experimental observables 
related to azimuthal anisotropic flow are found to be sensitive to shear viscous effects. The shear
viscosity decreases the anisotropy of the fluid velocity. Hence $v_{2}$ and $v_{4}$ 
as a function of transverse momentum ($p_{T}$) are expected to decrease with the increase in the value of $\eta/s$.

One of the main uncertainties in the estimation of $\eta/s$ using a viscous hydrodynamics simulation
is due to the choice of the initial conditions \cite{Chaudhuri:2009uk,Luzum:2008cw,Song:2010mg}. 
In this work we have considered two models, Glauber and
Color Glass Condensate (CGC), to obtain the initial transverse energy density profile. For this study
we have considered a smooth initial condition, which does not vary event-by-event. Previous work
have shown that both the spatial and momentum anisotropy are expected to be larger for a CGC based
initial condition compared to Glauber model~\cite{Luzum:2008cw}. Hence for other similar conditions in the simulations,
the calculations with CGC based initial condition is expected to give higher values of $v_{2}$ 
compared to initialization based on a Glauber model. Earlier comparisons of
viscous hydrodynamic simulations with both CGC and Glauber initial conditions to the experimental data 
at RHIC can be found in
Refs~\cite{Luzum:2008cw,Luzum:2009sb,Song:2010mg,Song:2011hk}. 
In Ref~\cite{Luzum:2008cw,Luzum:2009sb}, the experimental data used for comparison
are the centrality dependence of multiplicity, $\langle p_{T} \rangle$, $p_{T}$ integrated $v_{2}$, and minimum bias
$v_{2}$ vs. $p_{T}$ for charged hadrons in Au-Au collisions at $\sqrt{s_{NN}}$ = 200 GeV. In general it was 
observed that calculations with CGC based initial condition prefers a higher value of $\eta/s$ compared
to calculations with a Glauber based initial condition. In Ref~\cite{Song:2010mg,Song:2011hk} the authors have tried to
explain the centrality dependence of $v_{2}$ divided by the eccentricity with a viscous
hydrodynamic model for the QGP phase coupled to a transport model for the hadronic phase. 
Comparison of the experimental
data to the calculations done for CGC initial condition supports a $\eta/s$ value $\sim$ 0.16 - 0.24. While
the corresponding comparisons for a Glauber model based initial condition supports a lower value
of $\eta/s$ $\sim$ 0.08 - 0.16.

In the current work we compare the results from the viscous hydrodynamics simulations with two 
different initial conditions (Glauber and CGC) to recent high statistics measurements of 
$v_{2}(p_{T})$ and $v_{4}(p_{T})$ of charged hadrons in Au-Au collisions at $\sqrt{s_{NN}}$ = 200 GeV 
for a broad range in collision centrality (from 0-10\% to 40-50\%)~\cite{Adare:2010ux}. We also compare the simulated 
results to the measured charged particle invariant yields as a function of $p_{T}$ for various collision
centralities~\cite{Adler:2003au}.

The paper is organized as follows. In the next section we discuss the formalism of 
viscous hydrodynamic model used in this work. This includes a brief discussion on the 
energy-momentum conservation and relaxation equations for shear stress. We present a
detailed discussion on the initial conditions used in the calculations. The equation of
state used and the freeze-out conditions are also presented. In section III we present
a comparative study between calculations with Glauber and CGC initial conditions 
of various observables in the simulation. These includes the temporal evolution of 
shear stress, average transverse velocity, and  eccentricity. Section IV presents the
comparison of viscous hydrodynamic simulations with different input values of $\eta/s$
for both Glauber and CGC based initial conditions to the experimental data on invariant yield
versus $p_{T}$, $v_{2}(p_{T})$, and $v_{4}(p_{T})$ for various collision centralities. 
Finally in section V we present a summary of the work.

\section{Viscous hydrodynamic simulation}
In a relativistic viscous hydrodynamics scenario, there are two-fold corrections to 
the ideal fluid hydrodynamics. In presence of the dissipative processes, the energy momentum 
tensor contains additional dissipative corrections. The equilibrium freezeout distribution 
function used in the Cooper-Frey freezeout prescription ~\cite{Cooper:1974mv} also gets modified. 
The first order 
dissipative correction to the energy-momentum tensor leads to acausal behavior~\cite{Hiscock:1985zz}. The second order 
causal viscous hydrodynamics due to Israel-Stewart is one of the most commonly used theory \cite{Israel:1979wp}. 
For the simulation results presented here, we will follow the Israel-Stewart formalism 
for the evolution of a viscous fluid using the 2+1D viscous hydrodynamic code ``AZHYDRO-KOLKATA''~\cite{Chaudhuri:2008sj,Roy:2011pk}. 
Shear viscosity is the only dissipative process considered in our present study.
We assume a net-baryon free plasma is formed in Au-Au collisions at midrapidity at $\sqrt{s_{NN}}$ = 200 GeV.

\subsection{Conservation and relaxation equations}
The energy-momentum conservation equation and relaxation
equation for shear viscosity in Israel-Stewart formalism is
expressed as
\begin{eqnarray}  
\partial_\mu T^{\mu\nu} & = & 0,  \label{eq1} \\
D\pi^{\mu\nu} & = & -\frac{1}{\tau_\pi} (\pi^{\mu\nu}-2\eta \nabla^{<\mu} u^{\nu>}) \nonumber \\
&-&[u^\mu\pi^{\nu\lambda}+u^\nu\pi^{\mu\lambda}]Du_\lambda. \label{eq2}
\end{eqnarray}

Equation~\ref{eq1} is the conservation equation for the energy-momentum tensor, 
$T^{\mu\nu}=(\varepsilon+p)u^\mu u^\nu - pg^{\mu\nu}+\pi^{\mu\nu}$. 
$\varepsilon$, $p$, and $u$ are the energy density, pressure, and fluid
velocity respectively. $\pi^{\mu\nu}$ is the shear stress tensor. 
Equation~\ref{eq2} is the relaxation equation for the $\pi^{\mu\nu}$.   
$D=u^\mu \partial_\mu$ is the convective time derivative,
$\nabla^{<\mu} u^{\nu>}= \frac{1}{2}(\nabla^\mu u^\nu + \nabla^\nu u^\mu)-\frac{1}{3}  
(\partial_{\mu} u^{\mu}) (g^{\mu\nu}-u^\mu u^\nu)$ is a symmetric traceless tensor. 
$\eta$ is the shear viscosity and $\tau_\pi$ is the corresponding relaxation time.  
Assuming longitudinal boost-invariance, the above equations are solved  with the code 'AZHYDRO-KOLKATA'
~in ($\tau=\sqrt{t^{2}-z^{2}},x,y,\eta_{s}=\frac{1}{2}ln\frac{t+z}{t-z}$) coordinates.
Where $\tau$ is the longitudinal proper time, $(t, x, y, z)$ are space-time coordinates,
and $\eta_{s}$ is the space time rapidity.

\subsection{Initial conditions}

The initial conditions used here includes the initial energy density profile 
in the transverse plane ($\epsilon (x,y)$), the initial time
($\tau_{i}$), the transverse velocity profile $(v_{x}(x,y),v_{y}(x,y))$, shear stresses
in the transverse plane ($\pi^{\mu\nu}(x,y)$)  at $\tau_{i}$. The
$\tau_{i}$ value is taken as 0.6 fm. The
$\eta/s$ values are also inputs to the viscous hydrodynamics
simulations. We have taken the following temperature independent
values for this work, $\eta/s$ = 0, 0.08, 0.12, 0.16, and 0.18.

We have considered two different models for the calculation of 
initial energy density profile in the transverse plane. 
One is based on a two component Glauber model. At an impact
parameter $\bf{b}$, the transverse energy density is obtained from the following two 
component form

\begin{equation} 
\epsilon({\bf b},x,y) = \epsilon_0[(1-x_{h})\frac{N_{part}}{2} ({\bf b},x,y)+ x_{h} N_{coll}({\bf b},x,y)]
\label{eq:enprof}
\end{equation}  

\noindent where $N_{part}({\bf b},x,y)$ and $N_{coll}({\bf b},x,y)$ are the 
transverse profile of participant numbers and binary collision numbers 
respectively. $\epsilon_0$ corresponds to the central
energy density in $b$ = 0 and does not depend on
the impact parameter of the collision. The parameter $x_{h}$ is
the hard scattering fraction. Both $\epsilon_{0}$ and $x_{h}$ are fixed to
reproduce the experimental charged hadron multiplicity density at
midrapidity.  The $N_{part}({\bf b},x,y)$ and $N_{coll}({\bf b},x,y)$
values are obtained using an optical Glauber model calculation~\cite{Miller:2007ri}. 
The value of $x_{h}$ is found to be 0.9 and the values of
$\epsilon_{0}$ for various input values of $\eta/s$ are given in the Table~\ref{table1}.

\begin{table}[h]
\caption{\label{table1} Values of $\epsilon_{0}$ used in Glauber model
and normalization constant $C$ used in CGC model for initial transverse
energy density.} 
\begin{ruledtabular} 
\begin{tabular}{|c|c|c|}
$\eta/s$ &  $\epsilon_0$ (GeV/$fm^{3}$), Glauber & $C$ (GeV/$fm^{1/3}$) CGC   \\\hline
0.0          &   43.4      &  0.11            \\\hline
0.08        &   36.5       &  0.095           \\ \hline
0.12        &   32.5       &  0.085            \\\hline
0.16        &   27.7       &  0.070           \\ \hline
0.18        &   25.4       &  0.065            \\
\end{tabular}\end{ruledtabular}  
\end{table} 

The other model commonly used to obtain initial conditions for 
hydrodynamics is the Color-Glass-Condensate (CGC) approach, 
based on the ideas of gluon saturation at high energies~\cite{McLerran:1993ka,McLerran:1993ni}. We have used  the
KLN (Kharzeev-Levin-Nardi) $k_T$-factorization approach \cite{Kharzeev:2002ei}, 
due to Drescher {\it et al.} \cite{Drescher:2006pi}. 

We follow references \cite{Dumitru:2007qr,Luzum:2008cw} and consider that the initial
energy density can be obtained from the gluon number density through 
the thermodynamic relation,  
\begin{equation}
 \epsilon(\tau_i,{\bf x}_T,b)={\rm C}\times \left[ \frac{dN_g}{d^2 {\bf x}_{T}dY}({\bf x}_T,b)\right] ^{4/3},
\label{edCGC}
\end{equation}
where $\frac{dN_g}{d^2 {\bf x}_{T}dY}$ is the gluon number density evaluated at
central rapidity $Y=0$ and  the overall normalization $C$ is a free
parameter. $C$ is fixed to reproduce the experimentally
measured charged particle multiplicity density at midrapidity. The
values of $C$ used in the simulations for different
input values of $\eta/s$ are given in Table~\ref{table1}.
The number density of gluons produced in a  collision of two nuclei
with mass number $A$ is given by
\begin{widetext}
\begin{eqnarray}
 \frac{dN_g}{d^2 {\bf x}_{T}dY} = {\cal N}
  \int \frac{d^2{\bf p}_T}{p^2_T}
  \int^{p_T} d^2 {\bf k}_T \;\alpha_s(k_T) \;
  \phi_A(x_1, ({\bf p}_T+{\bf k}_T)^2/4;{\bf x}_T)\;
              \phi_A(x_2, ({\bf p}_T - {\bf k}_T)^2/4;{\bf x}_T),
 \label{eq:ktfac}
\end{eqnarray}
\end{widetext}
where ${\bf p}_T$ and $Y$ are the transverse momentum and 
rapidity of the produced gluons, respectively.  
$x_{1,2} = p_T\times\exp(\pm Y)/\sqrt{s}$ is the momentum fraction 
of the colliding gluon ladders with $\sqrt{s}$ the center of mass 
collision energy and $\alpha_s(k_T)$ is the strong coupling constant 
at momentum scale $k_T \equiv \left| {\bf k}_T \right|$. $\cal{N}$ is
the normalization constant.
The unintegrated gluon distribution functions are taken as
\begin{equation}  \label{uGDF}
\phi (x,k^2_{T}; {\bf x}_{T}) =
\frac{1}{\alpha_s (Q^2_s)} \frac{Q^2_s}{\textrm{max}(Q^2_s,k^2_{T})}
\,P({\bf x}_{T})(1-x)^4~,
\end{equation}
$P({\bf x}_{T})$ is the probability of finding at least one nucleon at
transverse position ${\bf x}_{T}$ and is defined as $P({\bf x}_{T}) = 1-\left(1-\frac{\sigma T_A}{A}\right)^A$,
where $T_A$ is the thickness function and $\sigma$ is the
nucleon-nucleon cross section taken as 42 mb. The saturation scale at a given momentum fraction $x$ and transverse coordinate ${\bf x}_{T}$ is given by
 $ Q^2_{s}(x,{\bf x}_T) = 2\,{\rm GeV}^2\left(\frac{T_A({\bf x}_T)/P({\bf x}_T)}{1.53/{\rm fm}^2}\right) \left(\frac{0.01}{x}\right)^\lambda$.
 The growth speed is taken to be $\lambda = 0.28$.

Shear stresses $\pi^{\mu\nu}$ is initialized to their corresponding
Navier-Stokes estimates for the boost invariance velocity profile, 
$\pi^{xx}=\pi^{yy}=2\eta/3\tau_i$, $\pi^{xy}=0$ \cite{Chaudhuri:2008je}.
We have used $\tau_\pi=3\eta/4p$ (where $\eta$ and $p$ are the shear viscous coefficient
and pressure) in our simulation, which corresponds to the 
kinetic theory estimates of relaxation time for shear viscous stress for a 
relativistic Boltzmann gas \cite{Israel:1979wp}. The initial values of
$v_{x}(x,y)$ and $v_{y}(x,y)$ are taken to be zero.

\subsection{Equation of state}

In the present simulations we have used an equation of state with
cross-over transition at temperature $T_c$ = 175
MeV~\cite{Roy:2011xt}.   The low temperature phase of the EoS is modeled  
by hadronic resonance gas, containing all the resonances with $M_{res}
\leq$2.5 GeV. The high temperature phase is a parametrization of the 
recent lattice QCD calculation \cite{Borsanyi:2010cj}. Entropy density
of the two phases are joined at $T$ = $T_c$ = 175 MeV by a smooth step 
like function.  The thermodynamic variables pressure ($p$), energy
density ($\varepsilon$), entropy density ($s$)
etc. are  then calculated by using the standard
thermodynamic relations

\begin{eqnarray}
p\left(T\right)&=&\int^{T}_{0}s\left(T^{\prime}\right)dT^{\prime}\\
 \varepsilon\left(T\right)&=&Ts\left(T\right)-p\left(T\right).
\end{eqnarray}

\subsection{Freeze-out condition}
The hydrodynamic expansion of the hot and dense matter leads to 
cooling of the system. After some time the mean-free path
of the constituent becomes large/comparable to the system size.
The system can no longer maintain the local thermal equilibrium
and the momentum distribution of the particles remains unchanged
after that. This is called freezeout. We use the Cooper-Frey algorithm 
at the freezeout to calculate invariant yields of the hadrons~\cite{Cooper:1974mv} . 
The freezeout  temperature which is a free parameter in the 
hydrodynamics simulation is taken as 
$T_{f}$=130 MeV. The effect of different choices of freeze-out temperature
on charged hadron $p_{T}$ spectra and elliptic flow is discussed in 
appendix~\ref{Appendix1}.

As we have already pointed out there are twofold
correction to the ideal fluid in the presence of viscous effects.
The freezeout distribution function for a system slightly
away from local thermal equilibrium can be approximated as \cite{Chaudhuri:2008sj}
\begin{eqnarray}
f_{neq}(x,p)=f_{eq}(x,p)[1+\phi(x,p)],
\end{eqnarray}
where $\phi(x,p)<<1$ is the corresponding deviation from the equilibrium 
distribution function $f_{eq}(x,p)$. The non-equilibrium correction
$\phi(x,p)$ can be approximated in Grad's 14 moment method by a quadratic function 
of the four momentum $p^{\mu}$ in the following way \cite{Muronga:2004sf,Muronga:2006zx}
\begin{eqnarray}
\phi(x,p)=\varepsilon-\varepsilon_{\mu}p^{\mu}+\varepsilon_{\mu\nu}p^{\mu}p^{\nu},
\end{eqnarray}
where $\varepsilon$, $\varepsilon_{\mu}$, and $\varepsilon_{\mu\nu}$
are functions of $p^{\mu}$, metric tensor $g^{\mu\nu}$, and
thermodynamic variables.

For our study where only shear stresses are considered, $\phi(x,p)$ has the following form
\begin{eqnarray}
\phi(x,p)=\varepsilon_{\mu\nu}p^{\mu}p^{\nu},
\end{eqnarray}
where
\begin{eqnarray}
\varepsilon_{\mu\nu}=\frac{1}{2(\epsilon+p)T^{2}}\pi_{\mu\nu}.
\end{eqnarray}
As expected, the correction factor increases with increasing values
of shear stress $\pi_{\mu\nu}$ at freezeout. The correction term also 
depends on the particle momentum. 
The Cooper-Frey formula \cite{Cooper:1974mv} for a non equilibrium system is
\begin{widetext}
\begin{eqnarray}\nonumber
\frac{dN}{d^{2}p_{T}dy}&=&\frac{g}{(2\pi)^{3}}\int d\Sigma_{\mu}p^{\mu}f_{neq}(p^{\mu}u_{\mu},T), \\\nonumber
\label{eq:chap3_shear_cooper}
\end{eqnarray}
\end{widetext}
where $g$ is the degeneracy of the particle considered and 
$d\Sigma_{\mu}$ is the normal to the elemental freeze-out hypersurface.

\section{Glauber versus CGC initial condition}

\subsection{Space-time evolution}

\bef
\begin{center}
\includegraphics[scale=0.4]{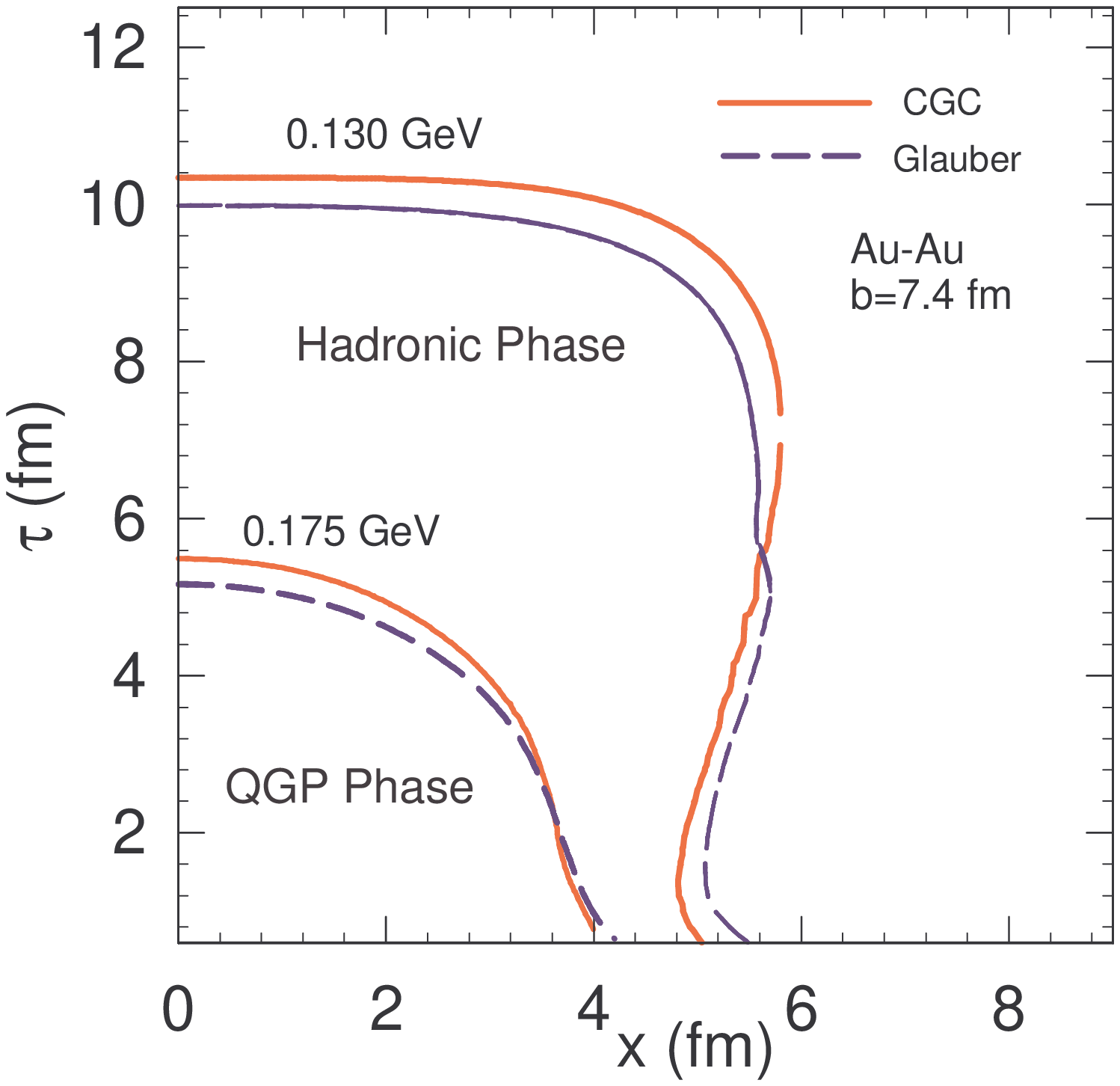}
\caption{(Color online) Constant temperature contours denoting the space time
boundaries of the QGP and hadronic phases from a 2+1D viscous
hydrodynamic simulation with $\eta/s$ = 0.08 for Au-Au
collisions at impact parameter 7.4 fm. The quark-hadron transition
temperature in the simulation is around 175 MeV and freeze-out
temperature is taken as 130 MeV. The solid red curves are simulations
with initial transverse energy density profile based on CGC model
while the dashed black curves correspond to initial conditions based
on Glauber model.}
\label{spacetime}
\end{center}
\eef
Figure~\ref{spacetime} shows the constant temperature contours
corresponding to $T_c$ = 175 MeV and $T_f$ = 130 MeV
in the $\tau$-$x$ plane (at $y$ = 0)  indicating the
boundaries for the QGP and hadronic phases respectively. The results
are from the viscous hydrodynamic simulations for Au-Au collisions at impact parameter 
7.4 fm and $\eta/s$ = 0.08. The solid red curves corresponds to
initial transverse energy density profile based on  CGC model and the 
dashed black curve corresponds to results based on Glauber model
initial conditions. We observe that the lifetime of QGP and hadronic
phases are slightly larger for the simulations based on CGC initial
conditions compared to Glauber based initial conditions.  The
spatial extent of the hadronic phase is slightly smaller for the
simulations with CGC initial conditions relative to Glauber based  
conditions.

\subsection{Temporal evolution of shear stress}

\bef
\begin{center}
\includegraphics[scale=0.4]{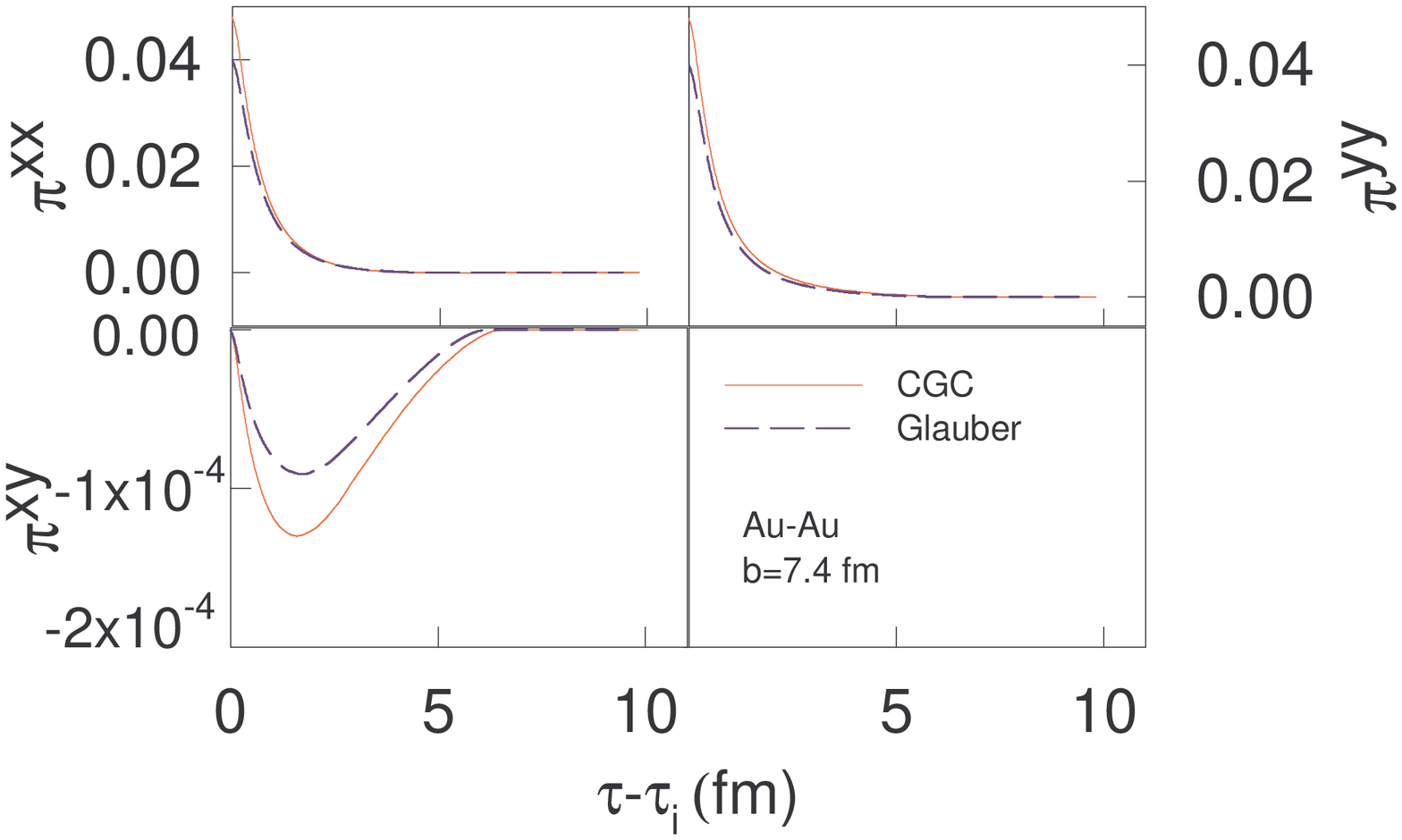}
\caption{(Color online) The spatially averaged  shear viscous stresses
  $\pi^{xx}$, $\pi^{yy}$, and $\pi^{xy}$ as a function of evolution
  time for Au-Au collisions at impact parameter 7.4 fm and $\eta/s$
  = 0.08. The solid red and black dashed curves corresponds to simulations with CGC
  and Glauber based initial conditions respectively.}
\label{avpi}
\end{center}
\eef

In presence of shear viscosity the thermodynamic pressure is modified.
The tracelessness of shear stress tensor $\pi^{\mu\nu}$, along with 
the assumption of longitudinal boost invariance ensures that at the initial time
$\pi^{xx}$ and $\pi^{yy}$ components of shear viscous stress are positive.
Consequently in viscous fluid the effective pressure in the transverse direction
is larger compared to the ideal fluid, for the same thermodynamic condition. 
It is then important to have some idea how various components of shear viscous 
stress $\pi^{\mu\nu}$ evolves in space-time. We have considered $\pi^{xx}$, $\pi^{yy}$, 
and $\pi^{xy}$  as the three independent components of shear stress $\pi^{\mu\nu}$. 
This choice is not unique.

The temporal evolution of spatially averaged $\pi^{xx}$, $\pi^{yy}$, and $\pi^{xy}$
are shown in Fig.~\ref{avpi} for CGC (solid red curve) and Glauber (black dashed curve) 
initialization of energy density. All the three components of $\pi^{\mu\nu}$ becomes zero after
time $\sim$ 7 fm irrespective of the CGC or Glauber model initialization.
At initial time the values of spatially averaged $\pi^{xx}$ and $\pi^{yy}$
are observed to be larger for CGC compared to the Glauber initialization. However, the
difference vanishes quickly $\sim$ 3 fm. For $\pi^{xy}$ a noticeable difference is seen 
for CGC and Glauber model initialization within time $\sim$ 6 fm.

\subsection{Average transverse velocity and Eccentricity}

\bef
\begin{center}
\includegraphics[scale=0.4]{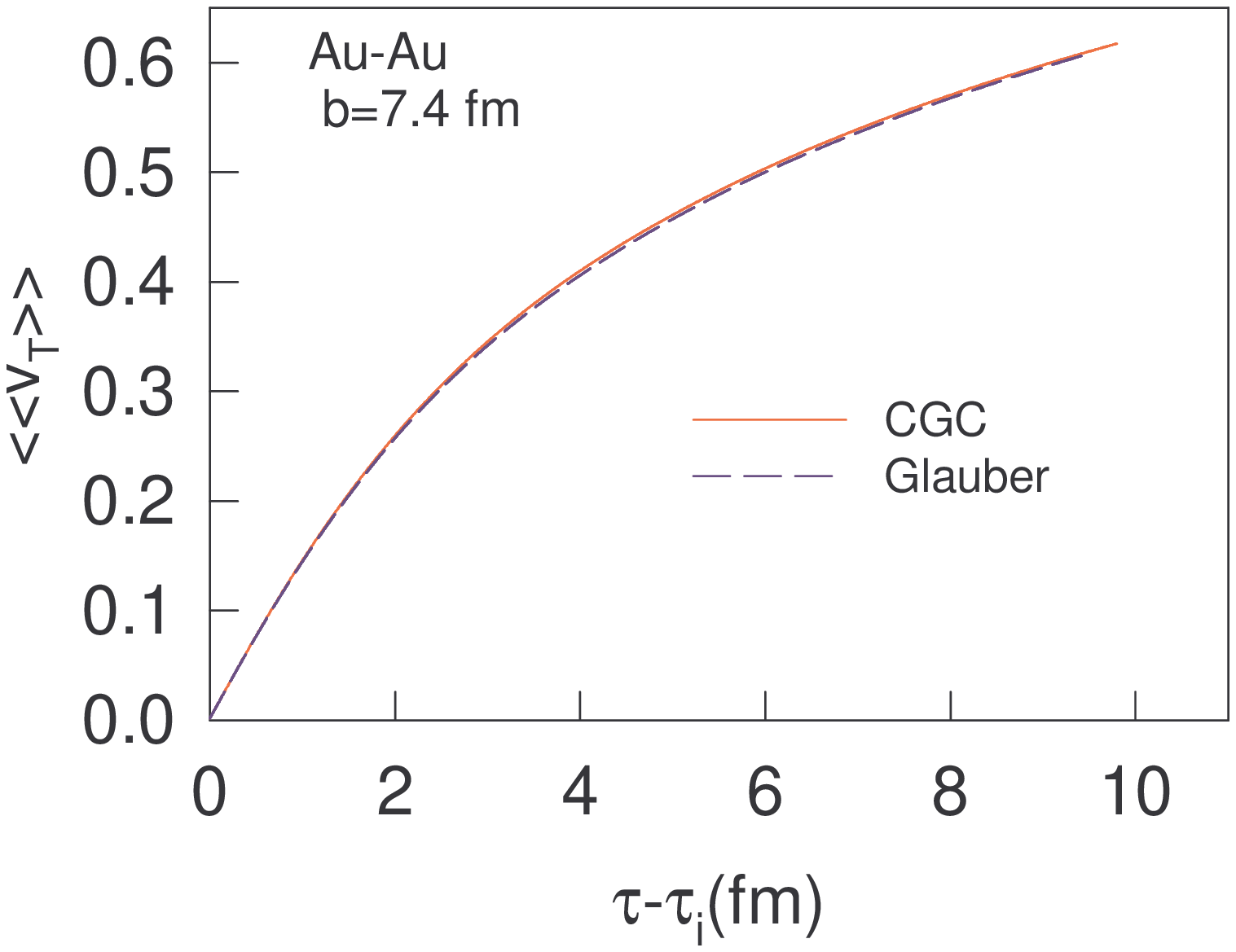}
\caption{(Color online) Time evolution of spatially averaged
  transverse velocity $\left\langle \left\langle v_{T}\right\rangle \right\rangle$. The results
  are from a 2+1D viscous hydrodynamic simulation with $\eta/s$
  = 0.08. The solid red curve corresponds to simulated result with
  CGC based initial transverse energy density profile. The black
  dashed line is the simulated result with Glauber based initial conditions.}
\label{transvel}
\end{center}
\eef
Figure~\ref{transvel} shows the temporal  evolution of the spatially 
averaged transverse velocity ($\left\langle \left\langle v_{T}\right\rangle\right\rangle$)
of the fluid with Glauber based and CGC based initial transverse
energy density profile with viscous hydrodynamic simulations carried
out for $\eta/s$ = 0.08. The simulation is done for Au-Au
collisions at impact parameter, b = 7.4 fm. The space averaged transverse velocity is defined as  
$\left\langle \left\langle
    v_{T}\right\rangle\right\rangle=\frac{\left\langle \left\langle
      \gamma\sqrt{v^{2}_{x}+v^{2}_{y}}~\right\rangle\right\rangle}{\left\langle
    \left\langle \gamma\right\rangle\right\rangle}$, where $\gamma =
\frac{1}{\sqrt{1-v_{x}^{2} - v_{y}^{2}}}$. 
The angular bracket $\left\langle \left\langle...\right\rangle\right\rangle$ 
implies an energy density weighted average.
Solid red curve is for CGC based initial condition and the dashed
black curve is for the Glauber based initial condition.
We observe almost no change in the 
$\left\langle \left\langle v_{T}\right\rangle\right\rangle$ as a
function of time for the two initial conditions studied. This effect
should be reflected in the slope of the invariant yield of the charged
hadrons as a function of transverse momentum being same for both the initial
conditions. These results are discussed in section IV A.

\bef
\begin{center}
\includegraphics[scale=0.4]{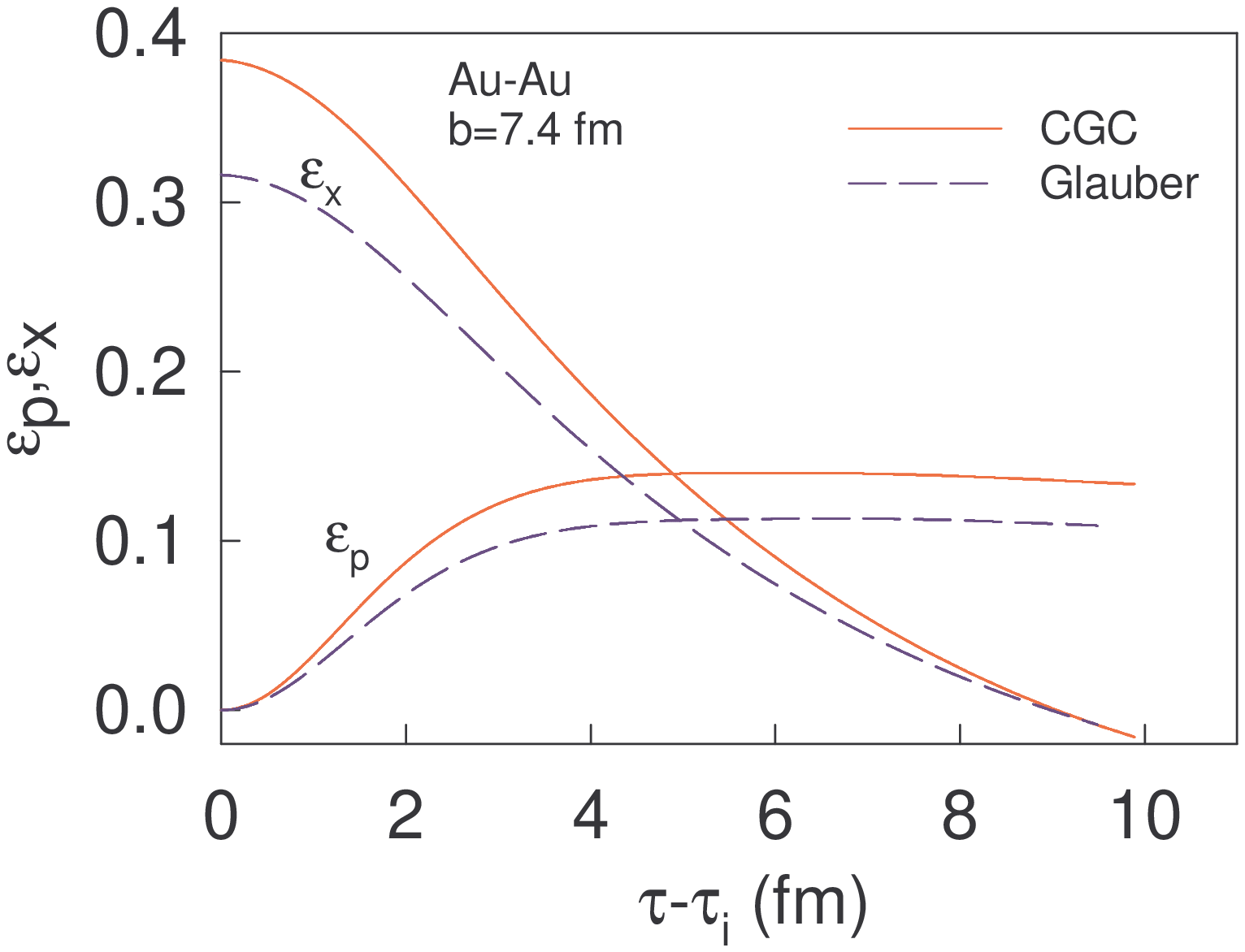}
\caption{(Color online) The temporal evolution of spatial
  ($\varepsilon_{x}$) and momentum ($\varepsilon_{p}$) eccentricity for Au-Au
  collisions at b=7.4 fm. The solid red curves corresponds to
  viscous hydrodynamics ($\eta/s$ = 0.08) simulated results with CGC
  based initial condition and the black dashed lines corresponds to
  results with Glauber based initial condition.}
\label{ecen}
\end{center}
\eef

Figure~\ref{ecen} shows the temporal evolution of the spatial eccentricity ($\varepsilon_{x}$)
and the  momentum space anisotropy ($\varepsilon_{p}$) of the viscous
fluid ($\eta/s$ = 0.08) with Glauber and CGC based initial conditions  for Au-Au
collisions at impact parameter, b = 7.4 fm.  The $\varepsilon_{x}$
which is a measure of the spatial deformation of the fireball from
spherical shape is defined as 
\begin{equation} \label{eq3}
	\varepsilon_{x}=\frac{\left\langle \left\langle y^{2}-x^{2}\right\rangle\right\rangle}{\left\langle \left\langle y^{2}+x^{2}\right\rangle\right\rangle},
\end{equation}
and the $\varepsilon_{p}$ which is a measure
of the asymmetry of fireball in momentum space is defined as  
\begin{equation}
	\varepsilon_{p}=\frac{\int dxdy (T^{xx}-T^{yy})}{\int dxdy (T^{xx}+T^{yy})},
 \end{equation}
where $T^{xx}$ and $T^{yy}$ are the components of energy-momentum tensor $T^{\mu\nu}$.
Solid red curve is for CGC based initial condition and the dashed
black curve is for the Glauber based initial condition. We find both
$\varepsilon_{x}$ and $\varepsilon_{p}$ are higher for the simulated
results with CGC based initial condition compared to initial
condition  based on Glauber model. As the simulated  elliptic flow
$v_{2}$ in hydrodynamic model is directly related to the temporal
evolution of the momentum anisotropy, we expect the $v_{2}$ for the CGC
based initial condition to be larger than the corresponding values for
the Glauber based initial condition. These results are discussed in
section IV B.

\section{Comparison to experimental data}

The experimental data used for comparison to our simulated results are
from the PHENIX collaboration at RHIC~\cite{Adare:2010ux,Adler:2003au}. The observables used are invariant
yield of charged hadrons, elliptic flow, and hexadecapole flow as a
function of $p_{T}$ for Au-Au collisions at pseudorapidity
$\mid \eta \mid < 0.35$ for $\sqrt{s_{\rm {NN}}}$ = 200 GeV. The high
statistics recent PHENIX measurement of 
elliptic ($k$ = 1) and hexadecapole ($k$ = 2) flow \cite{Adare:2010ux} are obtained using 
the formula $v_{2k} = \langle cos(2k (\phi - \Psi_{2})) \rangle$ after
correction of the event plane resolution. Where $\phi$ is the
azimuthal angle of the charged hadrons and $\Psi_{2}$ is the second
order event plane constructed using event plane detectors in $1.0 <
\mid \eta \mid < 3.9$. The rapidity gap between the detectors used to
measure the $v_{2k}$ and  $\Psi_{2}$  ensures absence of significant
$\Delta \eta$ dependent non-flow correlations, which are also absent
in our hydrodynamic simulations. $\Psi_{2}$ for our simulation is along $x$
axis. We compare below our simulated
results on invariant yield, elliptic, and hexadecapole flow for five
different collision centralities with input $\eta/s$ varying between
0.0 to 0.18 to the corresponding experimental data.
 
\subsection{Invariant yield}
\bef
\begin{center}
\includegraphics[scale=0.4]{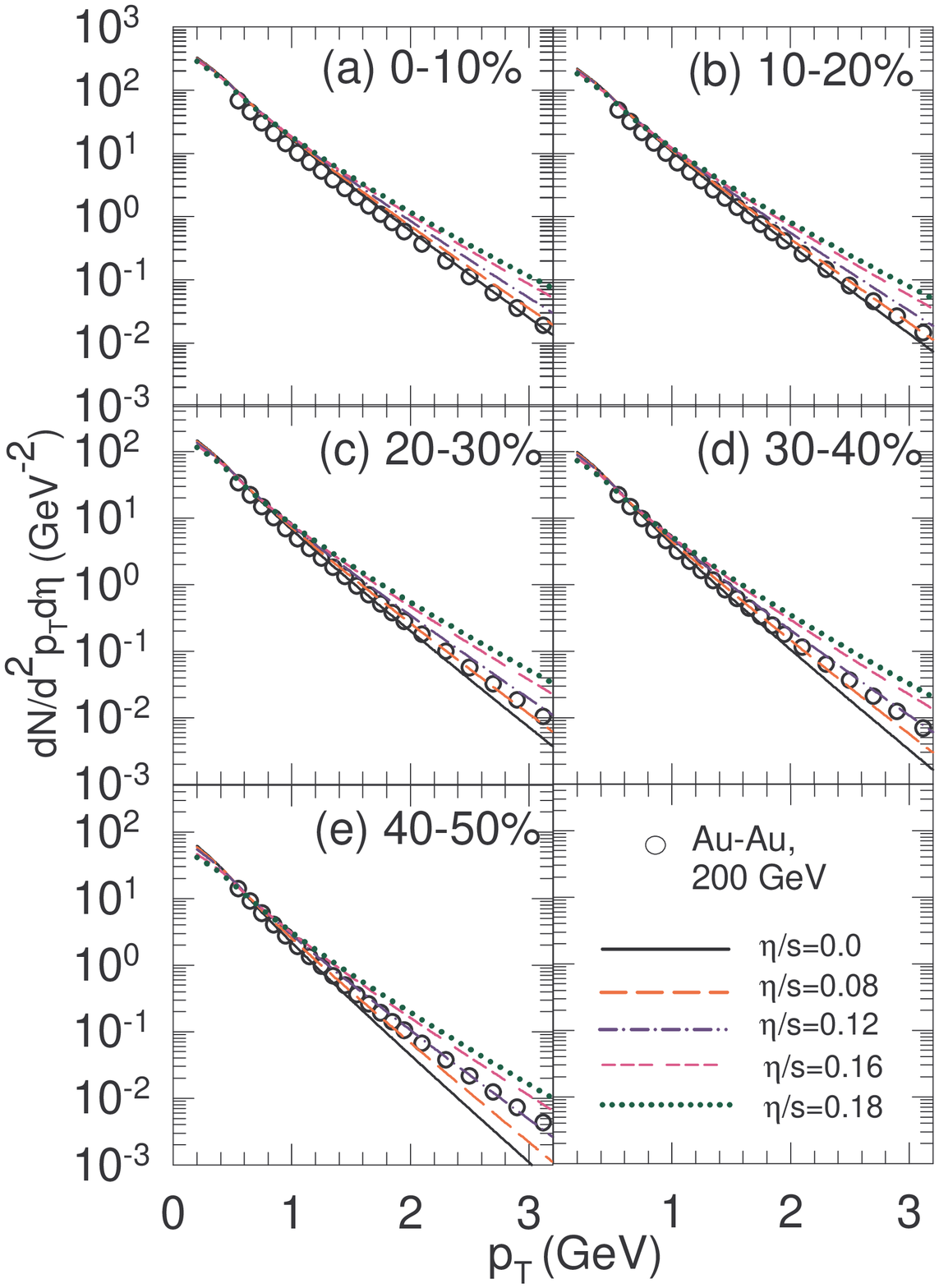}
\caption{(Color online) Invariant yield of charged hadrons as a
  function of transverse momentum at midrapidity for Au-Au collisions
  at $\sqrt{s_{\rm {NN}}}$ = 200 GeV. The open circles corresponds to
  experimental data measured by the PHENIX collaboration~\cite{Adler:2003au}. The lines
  represent results from a  2+1D relativistic
viscous hydrodynamic model with a Glauber based initial transverse
energy density profile and different $\eta/s$ values. The results are
shown for five different collision centralities 0-10\%, 10-20\%,
20-30\%, 30-40\%, and 40-50\%.}
\label{rhicspectraglauber}
\end{center}
\eef
Figure~\ref{rhicspectraglauber} shows invariant yield of charged
hadrons as a function of transverse momentum at midrapidity for Au-Au
collisions at $\sqrt{s_{\rm {NN}}}$ = 200 GeV for five different
collision centralities (0-10\%, 10-20\%, 20-30\%, 30-40\%, and
40-50\%). The open circles are the experimental data from the PHENIX
collaboration~\cite{Adler:2003au}. The simulated results are from the 2+1D relativistic
viscous hydrodynamic model with a Glauber based initial transverse
energy density profile. The black solid, orange long dashed, purple dash-dotted, magenta
short dashed and green dotted lines corresponds to calculations with
$\eta/s$ = 0.0, 0.08, 0.12, 0.16, and 0.18 respectively. We find the
0-10\% experimental data is best explained by simulation with $\eta/s$ =
0.0. Whereas data for collision centralities between 20-30\% to 40-50\%
supports a $\eta/s$ value within 0.08 to 0.12.

\bef
\begin{center}
\includegraphics[scale=0.4]{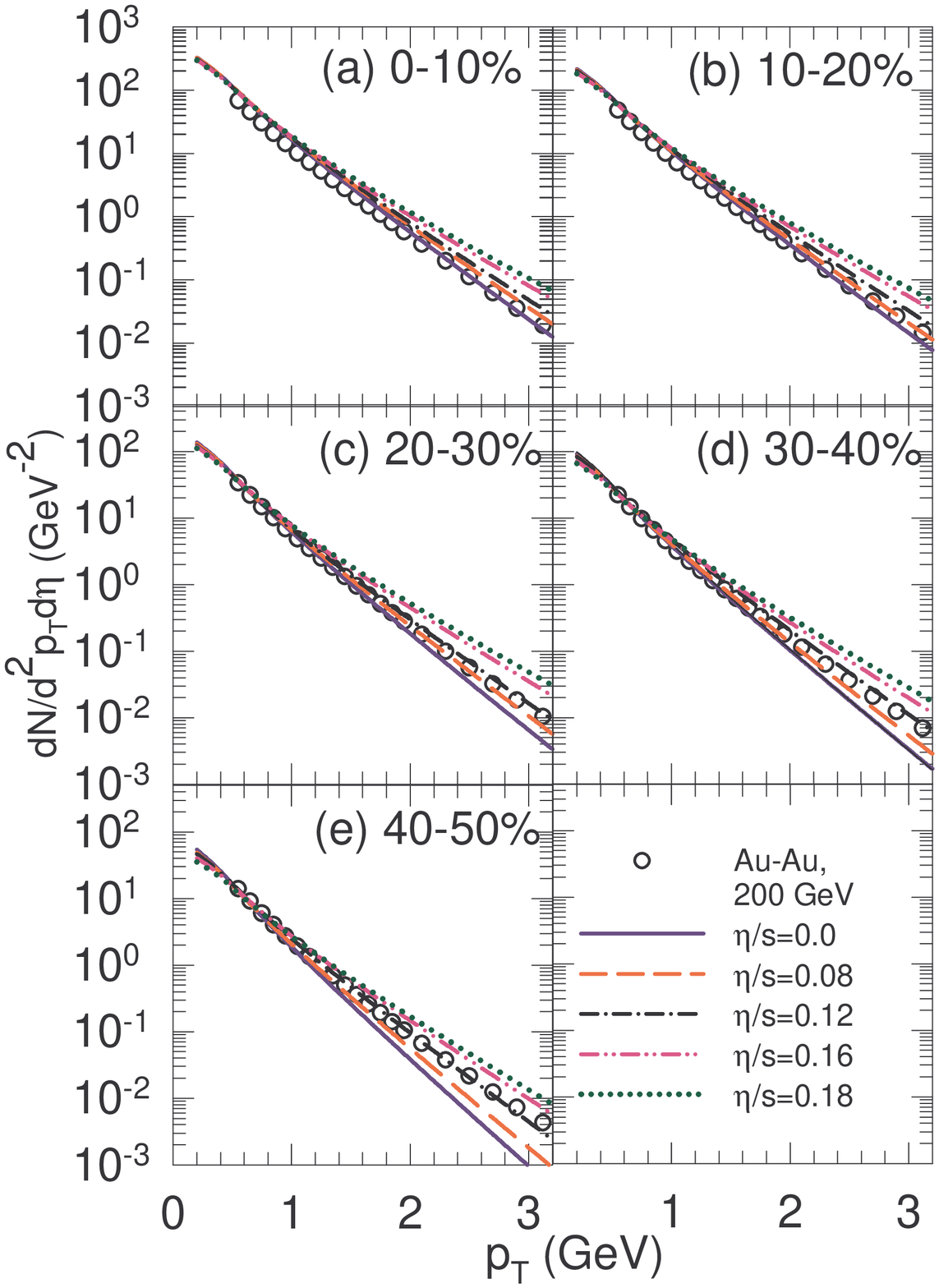}
\caption{(Color online) Same as Fig.~\ref{rhicspectraglauber} but the
  2+1D relativistic viscous hydrodynamic simulations are done with a
  CGC based initial transverse energy density profile.}
\label{rhicspectracgc}
\end{center}
\eef
Figure~\ref{rhicspectracgc} shows the same results as in
Fig.~\ref{rhicspectraglauber} but the simulated results corresponds to
2+1D viscous hydrodynamic calculations with a CGC based initial
transverse energy density profile. The conclusions regarding the
comparison between simulated results and experimental data are similar
to that obtained for Fig.~\ref{rhicspectraglauber}. This also means
that the invariant yield of charged hadrons are not very sensitive to
the choice of a Glauber based or CGC based initial conditions. The
average transverse velocity at the freeze-out which determines the
slope of the $p_{T}$ spectra was observed to be similar for the fluid
evolution with Glauber and CGC based initial conditions (see Fig.~\ref{transvel}).

\subsection{Elliptic flow}

\bef
\begin{center}
\includegraphics[scale=0.4]{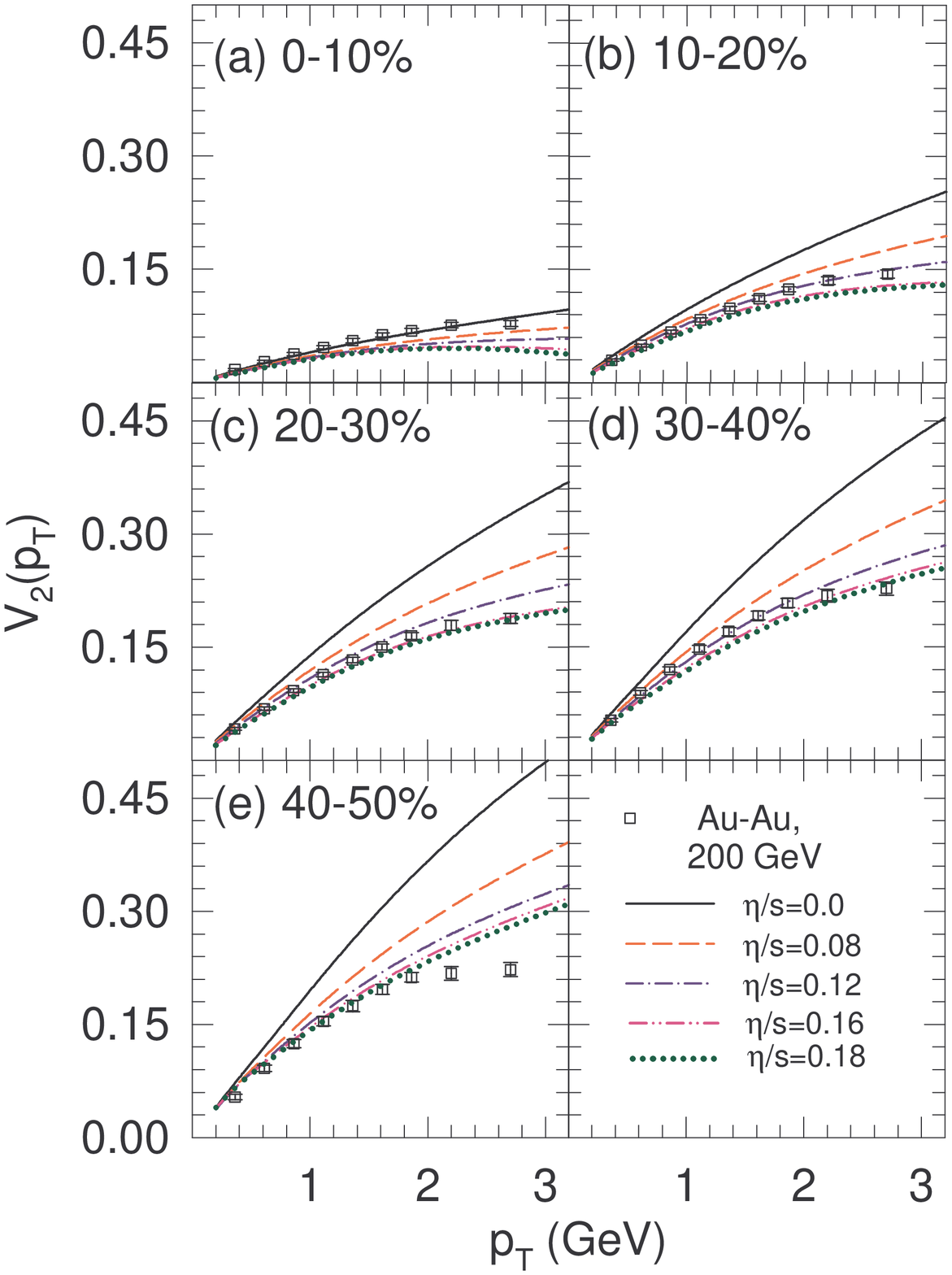}
\caption{(Color online) Elliptic flow of charged hadrons as a
  function of transverse momentum at midrapidity for Au-Au collisions
  at $\sqrt{s_{\rm {NN}}}$ = 200 GeV. The open circles corresponds to
  experimental data measured by the PHENIX collaboration~\cite{Adare:2010ux}. The lines
  represent results from a  2+1D relativistic
viscous hydrodynamic model with a Glauber based initial transverse
energy density profile and different $\eta/s$ values.}
\label{rhicv2glauber}
\end{center}
\eef
Figure~\ref{rhicv2glauber} shows the elliptic flow ($v_{2}$) as a
function of transverse momentum ($p_{T}$) for charged hadrons at
midrapidity in Au-Au collisions at $\sqrt{s_{\rm {NN}}}$ = 200 GeV.
The results are shown for five different
collision centralities (0-10\%, 10-20\%, 20-30\%, 30-40\%, and
40-50\%). The open circles are the experimental data from the PHENIX
collaboration~\cite{Adare:2010ux}. The simulated results are from the 2+1D relativistic
viscous hydrodynamic model with a Glauber based initial transverse
energy density profile. The black solid, orange long dashed, purple dash-dotted, magenta
short dashed, and green dotted lines corresponds to calculations with
$\eta/s$ = 0.0, 0.08, 0.12, 0.16, and 0.18 respectively. We find the
experimental data prefers higher values of $\eta/s$ as we go from
central to peripheral collisions. While 0-10\% collision centrality
experimental data is
best described by ideal fluid ($\eta/s$ = 0.0) simulation results,
those corresponding to 40-50\% collision centrality is closest to
simulated results with $\eta/s$ = 0.18.

\bef
\begin{center}
\includegraphics[scale=0.4]{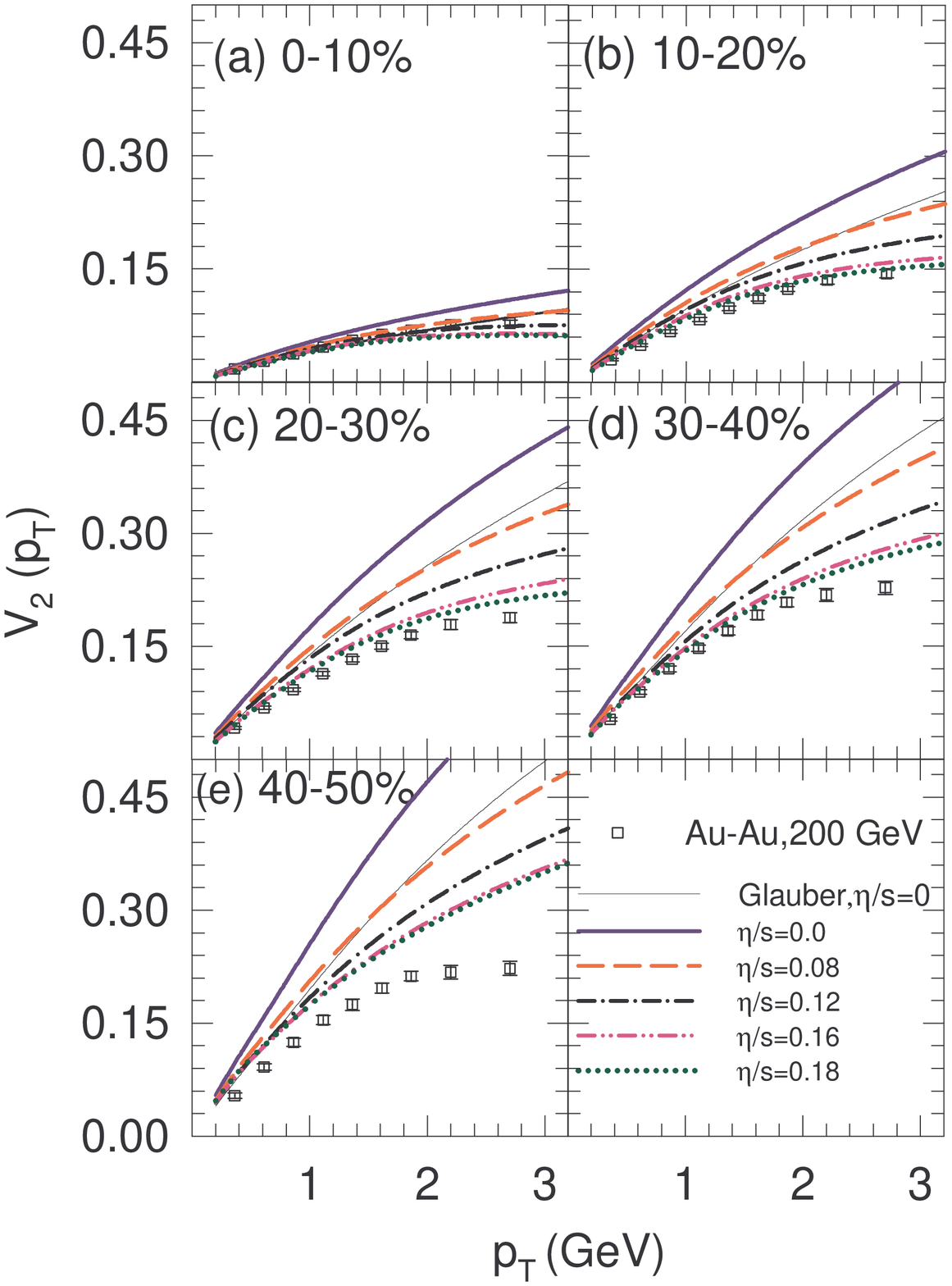}
\caption{(Color online) Same as Fig.~\ref{rhicv2glauber} but the
  2+1D relativistic viscous hydrodynamic simulations are done with a
  CGC based initial transverse energy density profile. Also shown for
  comparison is the result with Glauber based initial conditions for
  ideal fluid evolution.}
\label{rhicv2cgc}
\end{center}
\eef
Figure~\ref{rhicv2cgc} shows the same results as in
Fig.~\ref{rhicv2glauber} but the simulated results corresponds to
2+1D viscous hydrodynamic calculations with a CGC based initial
transverse energy density profile. Also shown for comparison the
simulated results for ideal fluid evolution with Glauber based initial
conditions. We find the $v_{2}(p_{T})$ for CGC based initial condition
is larger compared to corresponding results from Glauber based initial
conditions. This can be understood from the fact that CGC based
initial condition leads to a higher value of momentum anisotropy
compared to Glauber based initial condition (as seen in Fig.~\ref{ecen}). The
general conclusion that the experimental data prefers a higher value
of $\eta/s$ as we go from central to peripheral collisions as seen for
viscous hydrodynamic simulations with Glauber based initial conditions
also holds for those with the CGC based initial conditions. However, we
find from the comparison of experimental data to
simulations  based on CGC initial conditions that the $v_{2}(p_{T})$
data for 0-10\% collisions is best explained for simulated results
with $\eta/s$ between 0.08-0.12. This is in contrast to what we
saw from the comparisons of data to simulations with Glauber based
initial conditions, where the data preferred $\eta/s$ = 0.0 (see
Fig.~\ref{rhicv2glauber}).  For more peripheral collisions (centralities
beyond 20-30\%), it seems data would prefer a higher value of $\eta/s$
$\sim$ 0.18. We do not present simulation results for $\eta/s$ $>$
0.18 as the viscous hydrodynamic simulated spectra distributions show
a large deviations from ideal fluid simulation results (see appendix~\ref{Appendix2}).
This leads to a breakdown of the simulation frame work which is designed to be valid
for case of small deviations of observables from ideal fluid simulations.

\subsection{Hexadecapole flow}

\bef
\begin{center}
\includegraphics[scale=0.4]{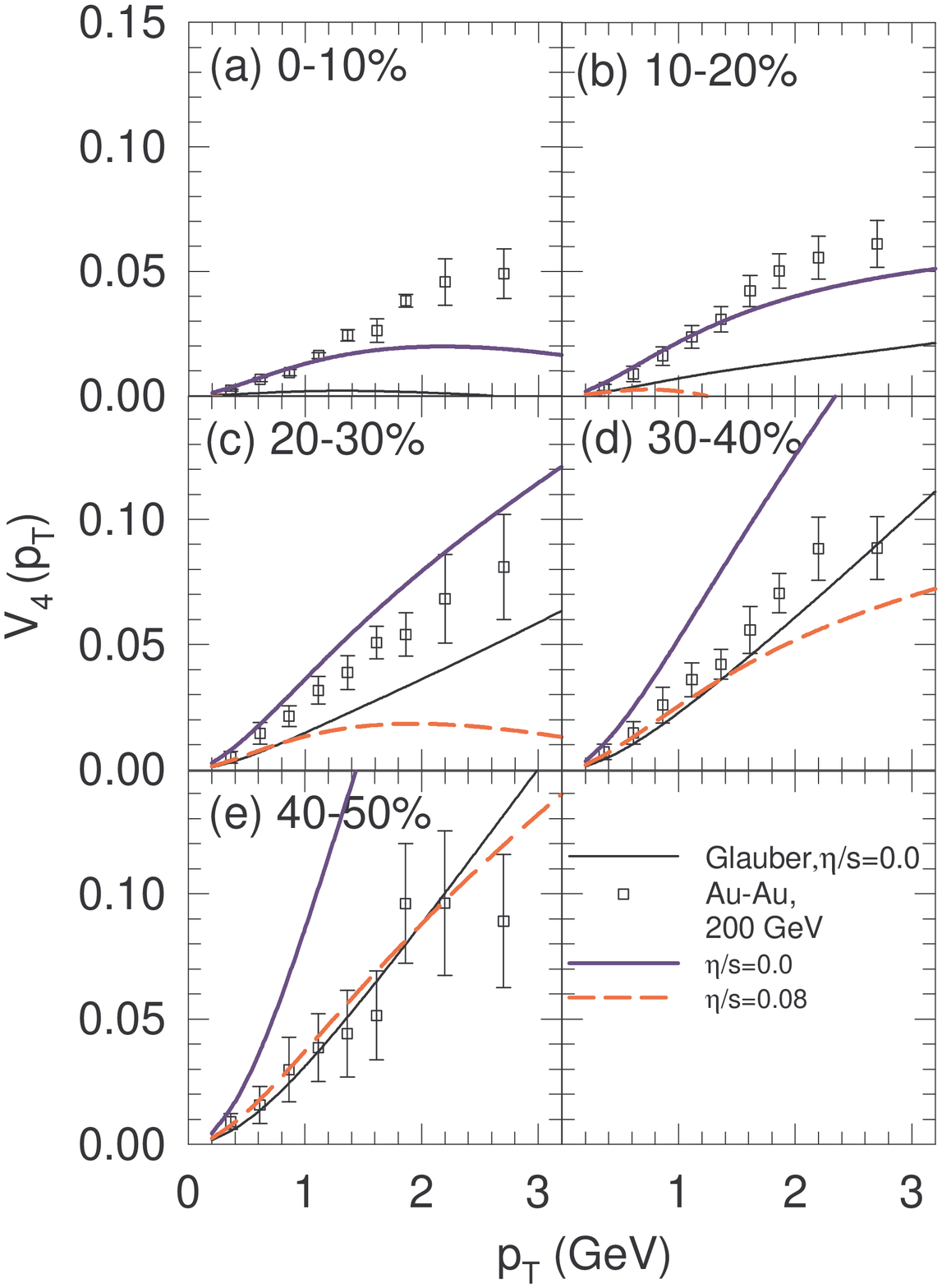}
\caption{(Color online) Hexadecapole flow of charged hadrons as a
  function of transverse momentum at midrapidity for Au-Au collisions
  at $\sqrt{s_{\rm {NN}}}$ = 200 GeV. The open circles corresponds to
  experimental data measured by the PHENIX collaboration~\cite{Adare:2010ux}. The curves
  represent results from a  2+1D relativistic
viscous hydrodynamic model with both Glauber based and CGC based  initial transverse
energy density profile and different $\eta/s$ values.}
\label{rhicv4}
\end{center}
\eef

Figure~\ref{rhicv4} shows the hexadecapole flow ($v_{4}$) as a
function of transverse momentum ($p_{T}$) for charged hadrons at
midrapidity in Au-Au collisions at $\sqrt{s_{\rm {NN}}}$ = 200 GeV.
The results are shown for five different
collision centralities (0-10\%, 10-20\%, 20-30\%, 30-40\%, and
40-50\%). The open circles are the experimental data from the PHENIX
collaboration~\cite{Adare:2010ux}. 
Simulated results for only ideal fluid evolution using
Glauber based initial condition are shown (solid black curve). While for the CGC based
initial conditions the simulated results are shown for $\eta/s$ = 0.0
(purple solid thick curve) and 0.08 (orange dashed curve). We do not
present simulated $v_{4}$  results for other $\eta/s$ values as these
are much lower compared to the data.  We find that $v_{4}(p_{T})$ from
ideal hydrodynamic simulations with Glauber based initial conditions
under predict the experimental data for all collision centralities
studied except for the most peripheral collisions (40-50\%) presented.
This is in sharp contrast to the observation for $v_{2}(p_{T})$ (see
Fig.~\ref{rhicv2glauber}) under similar conditions.  Comparsion
between simulated results with CGC based initial condition and
experimental data shows that the preferred $\eta/s$ lies between 0.0
and 0.08 for the collision centralities studied.
The $\eta/s$ values supported by the data on $v_{2}$ and $v_{4}$
using the simulated results presented here appears to be different.
In this study we have used smooth initial conditions, a more
realistic approach is to use a fluctuating initial condition 
and carry out event-by-event hydrodynamics. This will enable us to study 
the odd flow harmonics $v_{3}$ along with the even harmonics $v_{2}$ and $v_{4}$.
The simultaneous description of all these experimentally measured flow harmonics 
in a viscous hydrodynamics framework will probably provide a better estimation of $\eta/s$.

\section{SUMMARY}
We have carried out a 2+1D relativistic viscous hydrodynamic
simulation with two different initial conditions (Glauber and 
CGC) for the transverse energy density profile in Au-Au collisions
at $\sqrt{s_{NN}}$ = 200 GeV.  The simulations are carried out for
$\eta/s$ values between 0.0 to 0.18, using a lattice + hadron resonance
gas model based equation of state which has a cross over temperature for the
quark-hadron transition at 175 MeV. The shear viscous corrections are
considered both in the evolution equations and freeze-out distribution
function.

We find that the temporal dependence of the average
transverse velocity of the viscous fluid is similar for both the
initial conditions studied. The components
of shear viscous stress are observed to have higher values for the
simulations with CGC initial conditions compared to those for Glauber model
initialization at early times of fluid evolution ($<$ 6 fm).
The simulated invariant yield of charged
particles as a function of transverse momentum is also found to be
similar for the Glauber and CGC based initial conditions. The spatial
eccentricity and the momentum anisotropy have larger values for
simulations with CGC based initial condition compared to
the corresponding values for Glauber based initial condition.
The simulated elliptic flow is observed to be higher for calculations
with CGC based initial conditions relative to those with Glauber based
initial conditions, for a given collision centrality. 

We have compared our simulated results to the experimental data at
midrapidity on the centrality dependence  of invariant yield, $v_{2}$, and $v_{4}$  as 
a function of $p_{T}$ of charged hadrons measured in Au-Au  collisions at
$\sqrt{s_{\rm {NN}}}$ = 200 GeV.  From the comparison to the $p_{T}$
spectra of charged particles we observe that the data  supports a
$\eta/s$ value between 0 to 0.12 for the 0-10\% to 40-50\%
collision centralities for both the initial conditions considered.  The $v_{2}(p_{T})$ experimental data requires a
lower value of $\eta/s$ for simulations with Glauber model
initialization compared  to the CGC based initial conditions. For both the
models of initial conditions the $v_{2}(p_{T})$  data indicates a centrality
dependence in the estimated $\eta/s$ value, with peripheral collisions preferring larger
values. The experimental data on $v_{4}(p_{T})$ for the collision centralities
0-10\% to 40-50\% supports a $\eta/s$ value between 0 - 0.08 for a CGC
based initial condition. While simulated results using the Glauber
based initial condition for the ideal fluid evolution under estimates the $v_{4}(p_{T})$ for
collision centralities 0-10\% to 30-40\%.  Simulations with Glauber
model initial conditions explain the $v_{4}(p_{T})$ data for 40-50\%
collisions with $\eta/s$ = 0.0.  The observation associated with
$v_{4}$ is different from $v_{2}$, with $v_{4}$ data preferring smaller
values of $\eta/s$.  

There are further scopes of improvement on the simulated results
presented here. Recent experimental measurements of odd and higher
order azimuthal anisotropic flow~\cite{Adare:2011tg,ALICE:2011ab,Aad:2012bu} suggests that a fluctuating initial
condition needs to be considered. It is expected that the input
$\eta/s$ to the hydrodynamic simulations has a temperature dependence in
both the QGP and hadronic phases~\cite{Niemi:2011ix}. Although large uncertainties still
exist in the QCD computations of $\eta/s$ for the QGP phase. A more
precise estimation of $\eta/s$ would require the viscous fluid
simulations to also consider bulk viscosity and vorticity
effects~\cite{Becattini:2007sr}. Both of which are expected to be non-zero for the system
formed in high energy heavy-ion collisions and may affect the 
observables like $v_{2}$.  A proper prescription for bulk viscous
freeze-out correction is still under debate in literature~\cite{Roy:2011pk,Monnai:2009ad,Dusling:2011fd}, while
implementation of vorticity in viscous hydrodynamic
simulations have just started to be investigated. We plan to consider
some of these effects in the near future.

\noindent{\bf Acknowledgments}\\
BM is partially supported by
the DAE-BRNS project grant No. 2010/21/15-BRNS/2026.

\appendix
\section{Freezeout temperature}
\label{Appendix1}
\bef
\begin{center}
\includegraphics[scale=0.4]{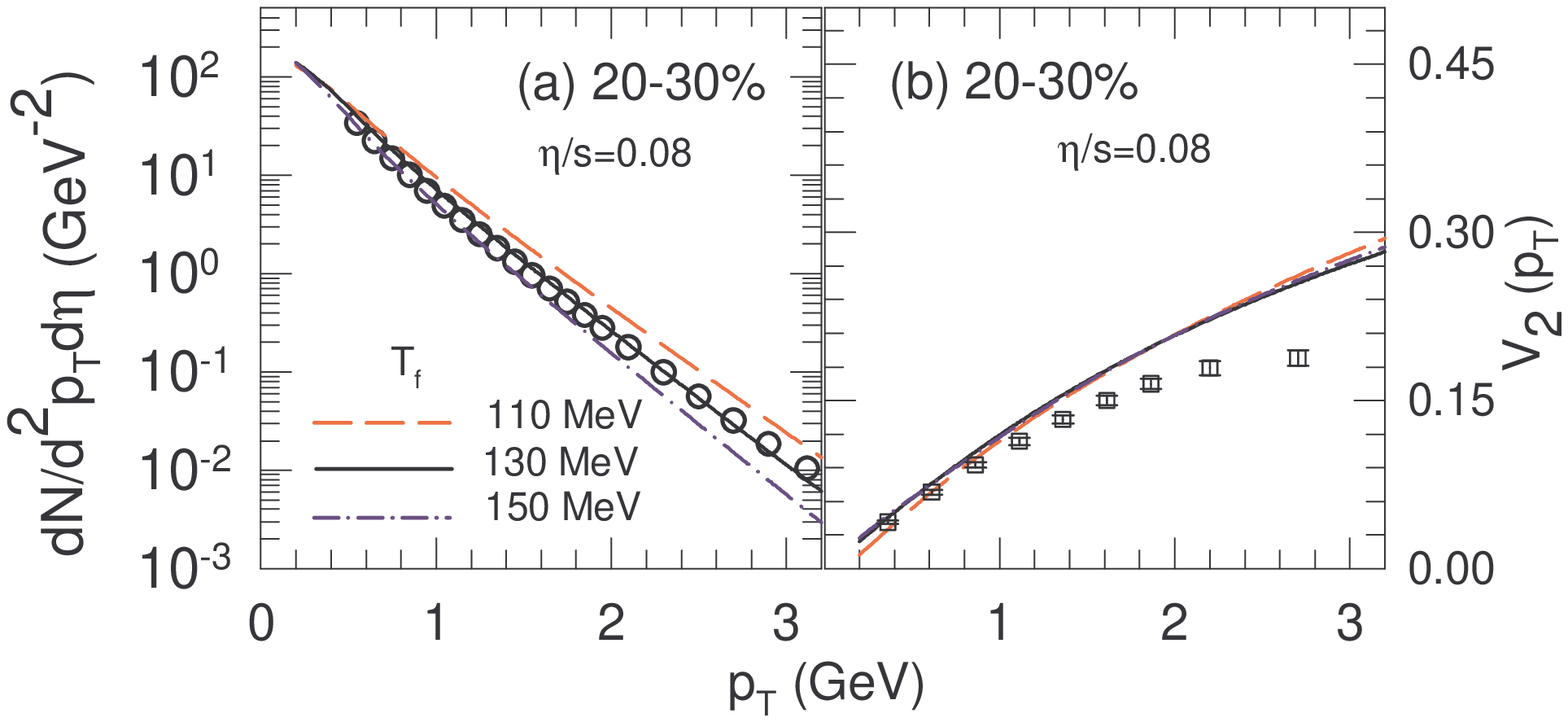}
\caption{(Color online) 
(a) Invariant yield and (b) elliptic flow of charged hadrons as a
  function of transverse momentum at midrapidity for Au-Au collisions
  at $\sqrt{s_{\rm {NN}}}$ = 200 GeV. The open circles corresponds to
  experimental data measured by the PHENIX 
  collaboration~\cite{Adare:2010ux}. The lines
  represent results from a  2+1D relativistic
viscous hydrodynamic model with a Glauber based initial transverse
energy density profile with $\eta/s$ = 0.08 and different $T_{f}$ values.}
\label{specv2tf}
\end{center}
\eef

We have studied the effect of different freeze-out temperature on 
the charged hadron invariant yield and elliptic flow. Figure~\ref{specv2tf}
shows the invariant yield (panel - a) and elliptic
flow (panel - b) of charged hadrons for 20-30\% centrality 
Au-Au collisions at $\sqrt{s_{\rm {NN}}}$ = 200 GeV. All the simulated 
results are for $\eta/s$ = 0.08 using the same Glauber based initial 
condition but with three different freeze-out temperatures, 
$T_{f}$ = 110 (red dashed curve),130 (black solid curve), and
150 MeV (blue dash-dotted curve). 

The slope of the $p_{T}$ spectra increases as $T_{f}$ decreases. 
This is because of higher radial velocity gained due to longer 
duration of evolution of the system for the lower freeze-out 
temperature. The experimentally measured $p_{T}$ spectra is 
best explained for simulations with input parameters as specified 
in section II and $T_{f}$ = 130 MeV.

For the current simulations the effect of different $T_{f}$ values
studied is observed to be small on elliptic flow of charged hadrons.
This could possibly be due to saturation of the value of the momentum
anisotropy at the early time of evolution.

\section{Viscous correction}
\label{Appendix2}
There are two kinds of dissipative correction to the ideal fluid simulation. First the energy momentum
tensor contains a viscous correction and the freezeout distribution function is also modified in presence of the dissipative processes. The viscous hydrodynamics model is applicable  when the dissipative correction
(both in the energy-momentum tensor and freezeout distribution function) is small  compared to the 
corresponding equilibrium value. It is then implied that the relative viscous correction ($\delta N/N_{eq}$) 
is small for the Israel-Stewart's hydrodynamics to be applicable, where $N_{eq}$ is the invariant yield
for system in local thermal equilibrium, and  $\delta N=\frac{dN}{d^{2}p_{T}}|_{viscous}-\frac{dN}{d^{2}p_{T}}|_{equilibrium}$. In figure \ref{speccorr}
the relative viscous correction $\delta N/N_{eq}$ for $\eta/s$ = 0.18 are shown for 0-10\% (black solid curve) and
40-50\% (black dashed curve) collision centrality. We observe that the relative shear viscous correction
is quite large ($\sim$ 50\%) at $p_{T}\sim$1.0 GeV for 40-50\% centrality (dashed curve in figure~\ref{speccorr}). 
The solid curve in the same figure shows the relative viscous correction to the invariant yield of charged hadron for 
Au-Au collision for 0-10\% centrality  at $\sqrt{s_{\rm {NN}}}$ = 200 GeV.
A higher value of $\eta/s$ will introduce a larger viscous correction and eventually the viscous hydrodynamics 
framework will no longer be applicable. 
\bef
\begin{center}
\includegraphics[scale=0.4]{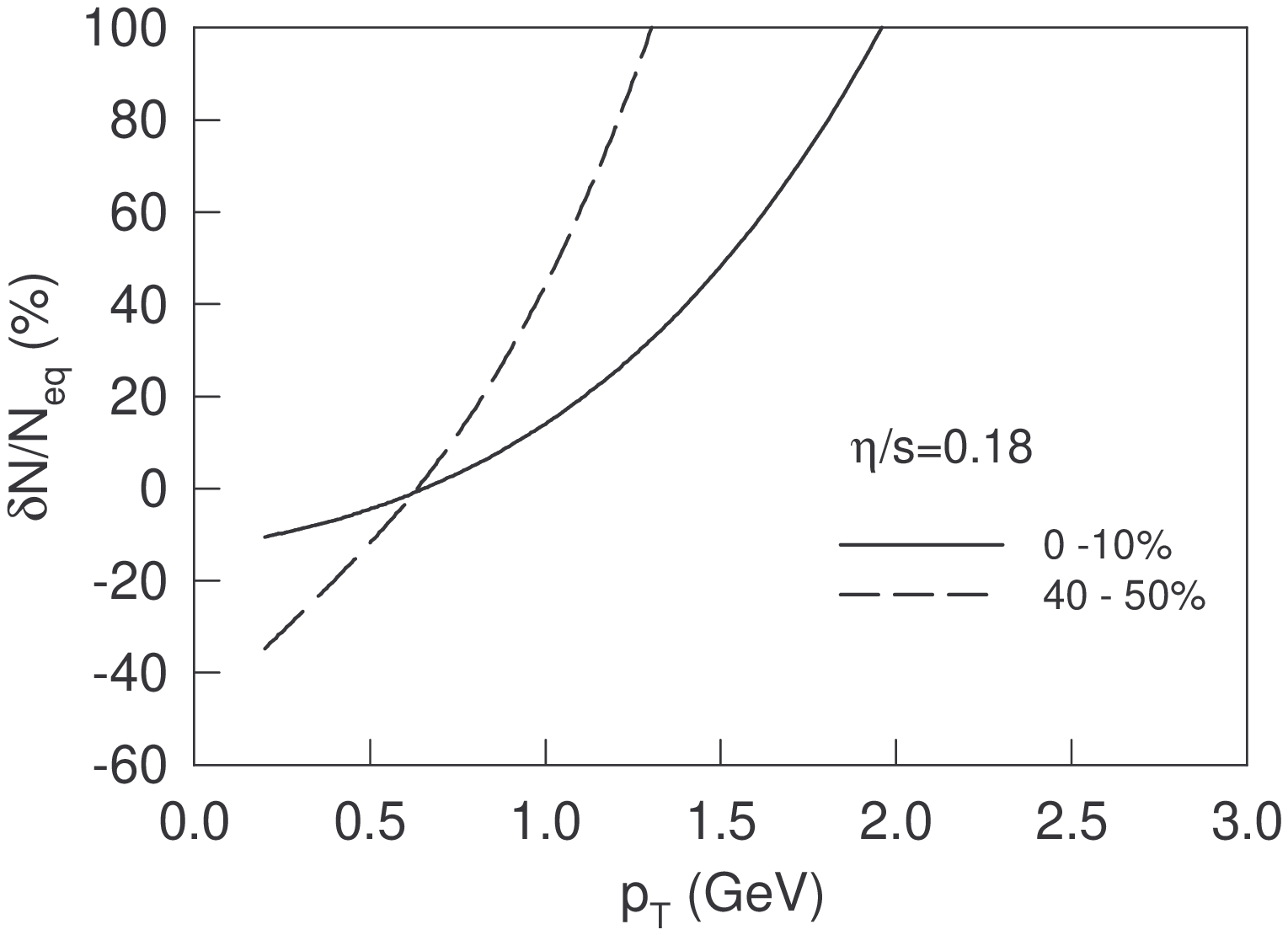}
\caption{ Dissipative correction to the invariant yield of the charged hadron as a function of 
$p_{T}$ for 0-10\% (black solid curve) and 40-50\% centrality (black dashed curve) Au-Au collisions
at $\sqrt{s_{\rm {NN}}}$ = 200 GeV. 
The results are shown for $\eta/s$=0.18 and Glauber based initial condition.}
\label{speccorr}
\end{center}
\eef

\end{document}